\documentclass[aps,prl,reprint,groupedaddress,showpacs]{revtex4-1}

\usepackage{indentfirst}
\usepackage{amsmath}
\usepackage{textcomp}
\usepackage{amssymb}
\usepackage{nicefrac}
\usepackage{marvosym}
\usepackage{yfonts}
\usepackage{notes2bib}
\usepackage{yfonts}
\usepackage{courier}
\usepackage{array}
\usepackage{graphicx}
\usepackage{nopageno}
\usepackage{txfonts}
\usepackage[usenames]{color}

\begin{document}
\title{Polaron-induced band renormalization due to linear and quadratic electron-phonon coupling}

\author{Pablo Garc\'ia Risue\~no}
\email[]{garcia.risueno@gmail.com}
\affiliation{Humboldt-Universit\"at zu Berlin, Physics Department and IRIS Adlershof,
Zum Gro\ss en Windkanal 6, 12489 Berlin, Germany}

\author{Dmitrii Nabok}
\affiliation{Humboldt-Universit\"at zu Berlin, Physics Department and IRIS Adlershof,
Zum Gro\ss en Windkanal 6, 12489 Berlin, Germany}

\author{Qiang Fu}
\affiliation{Humboldt-Universit\"at zu Berlin, Physics Department and IRIS Adlershof,
Zum Gro\ss en Windkanal 6, 12489 Berlin, Germany}

\author{Claudia Draxl}
\email[]{claudia.draxl@physik.hu-berlin.de}
\homepage[]{http://sol.physik.hu-berlin.de}
\affiliation{Humboldt-Universit\"at zu Berlin, Physics Department and IRIS Adlershof,
Zum Gro\ss en Windkanal 6, 12489 Berlin, Germany}

\date{\today}

\begin{abstract}
We present a novel approach to electron-lattice interaction beyond the linear-coupling regime. Based
on the solution of a Holstein-Peierls-type model, we derive explicit analytical expressions for the eigenvalue spectrum of the Hamiltonian, resulting in a narrowing of bands as a function of temperature. Our approach enables the intuitive interpretation in terms of quasiparticles, i.e. polaron bands and dressed-phonon frequencies. Being nonperturbative, the formalism also applies in the strong-coupling case. We apply it to the organic crystal naphthalene, with the coupling strengths obtained by {\it ab initio} calculations. 
\end{abstract}

\maketitle
The coupling between electronic and vibrational degrees of freedom is one of the most prominent examples of fermion--boson interaction in condensed-matter systems. Its theoretical description and analysis is central to the understanding of many physical phenomena like heat and charge transport, or superconductivity \cite{mahanbook,cohen1993}. The complex and possibly strong interaction between electrons and phonons makes perturbative approaches often problematic. A way to obtain non-perturbative solutions of the system Hamiltonian is to perform transformations which allow for a description of the system in terms of independent dressed particles instead of coupled bare particles. This is the concept behind the polaron transformation, which leads to analytical solutions in the linear-coupling regime \cite{mahanbook}. Analogous methods for the {\it nonlinear} nonlocal electron-phonon coupling (EPC) have not been available so far.

It is known from model Hamiltonians \cite{berciu1} and from applications of Green-function theory \cite{cardonaRMP,giustinoPRL2010,Marini2011,Marini2014,gonze2011,AntoniusPRL,ponceprb} as initiated in early work by Allen, Heine, and Cardona \cite{AH76,Allen1981,Allen1983} that the renormalization of electronic bands due to quadratic electron-phonon coupling is often of the same order as the linear one. Since these two contributions are often opposite in sign \cite{gonze2011}, the neglect of quadratic coupling can lead to significant errors in electronic-structure calculations considering vibrational effects. The relevance of quadratic EPC terms has also been pointed out for properly describing a number of phenomena, like band renormalization and spin crossover \cite{davino2011}, charge transfer \cite{girlando1989} or the Jahn-Teller effect \cite{koizumi2000}, and many materials like graphene \cite{lozovik2010,ziegler2011,basko2008}, SrTiO$_3$ \cite{liu2001}, molecular systems \cite{gonze2011} or systems with torsional motions \cite{coropceanu2007}. Recent research indicates dramatic effects of even small nonlinear EPC in model Hamiltonians \cite{Li2015,Li2015a}. Despite the prominent character of the quadratic electron-phonon interaction, just a few non-perturbative model approaches exist in the literature \cite{munnsilbey77,Riseborough1984,wohlman85,olsen2009,adolphsberciu,adolphs2014}.

In this work, we present a novel formalism for the {\it non-perturbative} treatment of {\it linear and quadratic} electron-phonon interaction. We derive an analytical solution that facilitates the understanding of the essential underlying physics in an intuitive way. It also allows for obtaining accurate solutions in the strong coupling regime, where a perturbative treatment fails, and for tackling large systems that would not be numerically affordable \cite{notta1}. Moreover, the \emph{nonlocality} of the Hamiltonian makes our procedure quite generally applicable, and, thus for instance, suited to describe phenomena like charge transport \cite{coropceanu2009,coropceanu2018,hanchargetransport}. 

Our theory is based upon a quadratic Holstein-Peierls-type model for the interaction between electrons and phonons. An analogous approach, including linear coupling only, has been successfully applied previously \cite{PRB1}. Using a tight-binding description for crystalline materials, the corresponding Hamiltonian takes the form
\begin{align}\label{ham11}
H \ & = \  H_{el} \ + \ H_{ph} \ + \ H_{el-ph}  \nonumber \\
 \ & = \ \sum_{m n }\ \varepsilon_{mn} \ a^{\dag}_m a_n  \ + \  \sum_{\bf{Q}} \ \hbar \omega_{\bf{Q}}  \ (b^{\dag}_{\bf{Q}}  b_{\bf{Q}}  +  \textrm{\textonehalf})   \\
 \  & \quad + \sum_{m n {\bf{Q}}}    \ \hbar \omega_{\bf{Q}} \ g_{{\bf{Q}} mn} \  a^{\dagger}_m a_n  \ (b^{\dagger}_{\bf{Q}}  + b_{-\bf{Q}} )   \nonumber \\
 \ & \quad + \ \sum_{m n \bf{Q}}  \ \hbar \omega_{\bf{Q}}  \ f_{{\bf{Q}} mn} \  a^{\dag}_m a_n \ (b^{\dag}_{\bf{Q}}  + b_{-\bf{Q}} ) (b^{\dag}_{-\bf{Q}}  + b_{\bf{Q}} ) \ . \nonumber
\end{align}
The electronic subsystem is described by $H_{el}$ where the fermionic operators $a_m $ ($a^{{\dag}}_m $) correspond to the annihilation (creation) of electrons at lattice site ${\bf R}_m$ with on-site energies $\varepsilon_{mm}$ and the strength of their interatomic coupling being governed solely by the transfer integrals $\varepsilon_{mn}$. The phononic subsystem is given by $H_{ph}$ where the bosonic operators $b_{\bf{Q}}$ ($b^{{\dag}}_{\bf{Q}}$) describe the annihilation (creation) of a phonon mode $\nu$ with wave vector $\bf{q}$ at frequency $\omega_{\bf{Q}}$ [$\bf{Q} \equiv (\bf{q},\nu)$]. The interaction between the electronic and vibrational degrees of freedom is governed by $H_{el-ph}$. The electron-phonon matrix elements $g_{{\bf{Q}}mn}$ and $f_{{\bf{Q}} mn}$ (note that $f_{{\bf{Q}}}$, $g_{{\bf{Q}}}$ are matrices) can be determined using the methods presented in Refs. \cite{giustinoPRL2010,gonze2011,ponceprb} or \cite{PRB1}, or by fitting to experimental data \cite{zhugayevych2015}. These coefficients correspond to the linear and quadratic electron-phonon coupling, respectively, because $ (b^{\dagger}_{\bf{Q}}  + b_{-\bf{Q}} )$ is proportional to the Fourier transform of the nuclear displacements \cite{SM}. We explicitly take into account both the local ($m=n$) Holstein-like terms as well as the nonlocal ($m \neq n$) Peierls-like terms, which may lead to remarkably different results than considering just the local ones \cite{SM,PRB1}.

Our approach relies on a unitary transformation that makes the Hamiltonian (Eq. \ref{ham11}) diagonal and thus facilitates an analytical solution. To achieve this goal in a non-perturbative manner, all the terms of the Hamiltonian that do not conserve the number of phonons are removed. By this procedure, we control the complexity that originates from the {\it nonlocal} Peierls-type coupling in $H_{el-ph}$. To this end, we adopt approximations that have been applied in a variant of the polaron transformation \cite{mahanbook,vanleubook} that has turned out successful in the linear-coupling case \cite{PRB1,JPCM}. This approach provides an analytical solution and allows for an intuitive interpretation within the quasiparticle picture. 

We propose the following canonical transformation of the Hamiltonian (Eq. \ref{ham11}):
\begin{subequations}\label{alphabetaHt}
\begin{align} 
& \tilde{H}  \,  =  \, e^S \,  H \,  e^{S^{\dag}{}} \   \label{SSS} \\
& S \  \equiv \ \sum_{m n}\ C_{   m n} \  a^{\dag}_m a_n    \\
& C_{mn}  \equiv   \sum_{\bf{Q}} \left[ \alpha_{{\bf{Q}} m n} \left(b^{\dagger}_{\bf{Q}} - b_{-\bf{Q}}\right) + 
  \beta_{{\bf{Q}}mn}  \ \left(b^{\dagger}_{\bf{Q}}b^{\dagger}_{-\bf{Q}} - b_{\bf{Q}}b_{-\bf{Q}}\right) \right]  
	\label{thav2} \\
& \alpha_{{\bf{Q}}mn} \equiv -\frac{1}{4} \left[\left( (\mathbb{I} + 4 f_{\bf{Q}})^{-1/4} - \mathbb{I} \right)^{-1}   (\mathbb{I} + 4 f_{\bf{Q}})^{-1} \textrm{ln}(\mathbb{I} + 4 f_{\bf{Q}})  g_{\bf{Q}} \right]_{mn} 
  \label{alphaHt} \\
 & \beta_{{\bf{Q}}  m n} \equiv \frac{1}{8} \big[\textrm{ln}(\mathbb{I} + 4 f_{\bf{Q}})\big]_{mn} 
  \label{betaHt}
\end{align}
\end{subequations}
In Eq. \ref{alphabetaHt}, the functions of matrices are represented by the corresponding Taylor expansion. By performing this transformation, the Hamiltonian (Eq. \ref{ham11}) becomes \cite{SM}:
\begin{equation}\label{ham31k}
\tilde{H}  =  \sum_{\bf{k}}  \ a^{\dag}_{\bf{k}} a_{\bf{k}} \ \left( \varepsilon_0 +
          \tilde{\varepsilon}'_{\bf{k}}  + \eta_{\bf{k}}  + \chi_{\bf{k}}  \right) 
             +   \sum_{\bf{Q}}  \hbar \omega_{\bf{Q}}
              \left( b^{\dag}_{\bf{Q}}  b_{\bf{Q}} + \textrm{\textonehalf}  \right) 
\end{equation}
where
\begin{subequations}\label{varsec3}
\begin{align}
\tilde{\varepsilon}'_{mn}  & \equiv \ {\big( e^C \ \varepsilon' \ e^{-C} \big)}_{mn} \ ; \  \varepsilon'_{mn}=\varepsilon_{mn} \textrm{ if } m\neq n \ , \ \ \varepsilon'_{mm}=0 \label{varepstilde} \\
\eta_{mn} & \equiv  \ \sum_{\bf{Q}} \ \hbar \omega_{\bf{Q}} \ 
                    \left( -  g_{\bf{Q}} \big( \mathbb{I} + 4 f_{\bf{Q}} \big)^{-1} g_{-\bf{Q}} \ \right)_{mn} \\
 \chi_{mn} & \equiv  \sum_{\bf{Q}} \ \hbar \omega_{\bf{Q}} \   \lambda_{{\bf{Q}} mn}
  \,  \left( \,  b^{\dag}_{\bf{Q}}  b_{\bf{Q}} \ + \ \nicefrac{1}{2} \right) \\
          \lambda_{{\bf{Q}} mn} &  \equiv  \left( \sqrt{ \mathbb{I} + 4 f_{\bf{Q}} } -  \mathbb{I} \right)_{mn} \\
 \Lambda_{\bf{k}} &  \equiv \frac{1}{N} \sum_{m n} \Lambda_{mn}  \ e^{i {\bf{k}} ({\bf{R}}_m - {\bf{R}}_n)} \quad \textrm{for }
 \Lambda = \tilde{\varepsilon}', \ \lambda_{{\bf{Q}} }, \ \eta, \ \chi \\
 a_{\bf{k}} & \equiv \frac{1}{\sqrt{N}} \sum_{{n}}  e^{ - i {\bf{k}} {\bf{R}}_n} a_n 
\end{align}
\end{subequations}
$N$ is the number of unit cells, and $\varepsilon_0$ is the value of the diagonal entries of the matrix of transfer integrals ($\varepsilon_0 \equiv \varepsilon_{mm}$) \cite{PRB1}. The different terms of $\tilde{H}$ (Eq. \ref{ham31k}) correspond to dressed electrons and phonons. $a^{\dag}_{\bf k} a_{\bf k}$ and $b^{\dag}_{\bf Q} b_{\bf Q}$ represent the number operators of electrons and phonons, respectively. 
Since $H$ (Eq. \ref{ham11}) contains terms which are not proportional to $b^{\dag}_{\bf Q} b_{\bf Q}$, {it is not straightforward to assign particle numbers to the eigenvalues of $H$. Conversely, our approach characterizes every eigenvalue of $\tilde{H}$ (Eq. \ref{ham31k}) after the evaluation of $\langle \tilde{\varepsilon}'_{\bf{k}} \rangle$ by the number of dressed electrons and phonons, respectively.} (A more detailed explanation of the meaning of the terms of (Eq. \ref{ham31k}) is presented in the Supplemental Material.) {With this evaluation, the transformed Hamiltonian $\tilde{H}$ (Eq. \ref{ham31k}) becomes diagonal, which is explained below (see Eq. \ref{analitthav}).} The advantage of our transformation (Eq. \ref{SSS}) becomes clear by noticing what follows: $H$ contains terms proportional to $( b^{\dag}_{\bf{Q}}b^{\dag}_{-\bf{Q}}{-}b_{\bf{Q}} b_{-\bf{Q}} )$ and to $( b^{\dag}_{\bf{Q}}{+} b_{-\bf{Q}} )$, which do not conserve the number of phonons. The accurate calculation of the eigenvalues of the Holstein-Peierls Hamiltonian by numerical means requires very large basis sets, which makes such calculation numerically very demanding \cite{zhugayevych2015}. Indeed, as shown in the Supplemental Material \cite{SM} for a 2D square lattice \cite{Li2015,Li2015a}, the numerical solution of this model system, including local and nonlocal electron-vibrational couplings, requires high phonon populations in the basis sets. In contrast, our analytical solver presented here provides accurate solutions very efficiently.

We proceed now by evaluating Eq. \ref{varepstilde}, which will allow us to quantify the polaron transfer integrals and to make $\tilde{H}$ diagonal. If the matrices $f_{{\bf{Q}}}$, $g_{{\bf{Q}}}$ commute with $\varepsilon'$, this leads to $\tilde{\varepsilon}'_{mn} = \varepsilon'_{mn}$ making $\tilde{H}$ diagonal without any approximation. If they do not commute, we can make $\tilde{H}$ diagonal by replacing $ \tilde{\varepsilon}'_{mn}$ by its thermal average $\langle\tilde{\varepsilon}'_{mn} \rangle$. Such an approach was employed to solve analogous problems for the linear \cite{PRB1,JPCM,zhugayevych2015} and quadratic cases \cite{munnsilbey77,wohlman85}. By proceeding in this way, an analytical solution becomes feasible, while the behavior of the system is still reliably reproduced. We note that the approximation of thermal averaging is commonly used in literature \cite{AH76,ueda2007,tamura2012,li2012prb,defilippis2015}. In our case it is particularly harmless \cite{PRB1,JPCM}, because the equations depending on phonon operators are proportional to commutators of $\varepsilon'$ the matrices $\alpha_{\bf{Q}}$ and $\beta_{\bf{Q}}$  \cite{SM}, which are expected to be very small or zero. The thermal averaging can also be avoided if applying perturbation theory on top of $\tilde{H}$, as done in Refs. \cite{munnsilbey77,wohlman85}.

For the explicit evaluation of the thermal averages $\langle \tilde{\varepsilon}'_{mn} \rangle$, we apply the {Baker-Campbell-Hausdorff (BCH) theorem \cite{PRB1,bchformula} to Eq. \ref{varepstilde}. This yields an infinite series that can be evaluated by truncating it at a finite order \cite{SM}. Nonetheless, an analytical expression for the entire series is also possible, leading to {an expression for the band narrowing as a function of temperature}
\begin{align}\label{analitthav}
& \langle \ \tilde{\varepsilon}'_{mn}  \ \rangle \  = \ \varepsilon'_{mn} \
 \  \textrm{exp} \Big[ - { \sum_{\bf{Q}} }^{\prime}   \mathbb{G}_{{\bf{Q}} m n}  \ (\nicefrac{1}{2} + N_{\bf{Q}})      \\ 
   & \qquad  \ \ \ \qquad \quad \qquad \
     - { \sum_{\bf{Q}} }^{\prime}  \mathbb{F}_{{\bf{Q}} m n}  \ ( (1 + N_{\bf{Q}})^2 + N^2_{\bf{Q}} ) \ \ \Big] \ , \nonumber \\
& \mathbb{G}_{{\bf{Q}} m n} \   \equiv \ | \ \alpha_{{\bf{Q}} m m} - \alpha_{{\bf{Q}} n n}  \ |^2 \ + \ 
             \sum_{l \neq m,n} \ \big( \ | \, \alpha_{{\bf{Q}} m l} \,|^2 + | \, \alpha_{{\bf{Q}} n l}  \, |^2 \ \big) \, ,   \nonumber \\
& \mathbb{F}_{{\bf{Q}} m n} \   \equiv \ | \ \beta_{{\bf{Q}} m m} - \beta_{{\bf{Q}} n n}  \ |^2 \ + \ 
             \sum_{l \neq m,n} \ \big( \ | \, \beta_{{\bf{Q}} m l} \,|^2 + | \, \beta_{{\bf{Q}} n l}  \, |^2 \ \big) \, ,   \nonumber
\end{align}
where $N_{\bf{Q}}$ is the Bose distribution and $^{\prime}$ in the summations means that one must omit the contributions of the ${\bf{Q}}$'s which satisfy $[f_{\bf{Q}},\varepsilon]=0$ (for $\mathbb{F}_{\bf{Q}} $), and $[f_{\bf{Q}},\varepsilon]=[g_{\bf{Q}},\varepsilon]=0$ (for $\mathbb{G}_{\bf{Q}} $) (this happens, e.g., if ${\bf{q}}=0$). In the derivation of (Eq. \ref{analitthav}) \cite{SM}, we have used the approximation $[g_{\bf{Q}}, g_{\bf{Q}'}]=[g_{\bf{Q}}, f_{\bf{Q}'}]=[f_{\bf{Q}}, f_{\bf{Q}'}]=0$  for all $g_{\bf{Q}}, g_{\bf{Q}'}$, and we only consider the most important contributions ($\varepsilon_{mn}$, $\varepsilon_{nm}$ and $\varepsilon_{mm}$) in the evaluation of the commutators of $f_{\bf Q}$ and $g_{\bf Q}$ with $\varepsilon$. Proceeding this way, corresponds to a generalization of approximations that where shown to have a very small effect \cite{PRB1,JPCM}. The terms $(1+N_{{\bf{Q}}})$ and ($\nicefrac{1}{2}+N_{\bf{Q}})$ in Eq. \ref{analitthav}} can be identified to correspond to the emission and absorption of \emph{one} phonon, respectively. They arise from the linear EPC. Analogously, the terms in $(1{+}N_{{\bf{Q}}})^2$ and $N_{{\bf{Q}}}^2$ correspond to emission and absorption processes of {\it two} phonons, which arise from the quadratic EPC term of $H$. The evaluation of $\langle \tilde{\varepsilon}'_{mn}  \rangle $, either using Eq. \ref{analitthav} or by truncating the BCH expansion of $\varepsilon'_{mn}$ \cite{SM}, makes the transformed Hamiltonian $\tilde{H}$ diagonal and provides analytical (symbolic) expressions for its eigenvalue spectrum. 

We illustrate the strength of the formalism presented above by applying it to the molecular organic semiconductor naphthalene, which crystallizes in a monoclinic structure (space group $P2_{1/a}$) with two non-equivalent molecules in the unit cell. It has been subject of previous investigations to demonstrate band-narrowing based on a tight-binding model, but including linear coupling only \cite{PRB1}. 
All the \textit{ab initio} calculations are performed with the VASP code \cite{vasppaper,vasppaper2}, that implements the projector augmented wave (PAW) method. Hard pseudopotentials \cite{ref3qiang,ref4qiang} are used for C and H atoms; exchange-correlation effects are described with the local-density approximation \cite{ldaperdewzunger}. A planewave cutoff energy of 1200 eV and a $6 \times 8 \times 6$ ${{\bf k}}$-mesh have been employed in the ground-state calculation. We adopt the experimental lattice parameters (at 5K) \cite{naphtexperimental} of a=8.08 \AA, b=5.93 \AA, c=8.63 \AA, and $\beta=124.7^{\circ}$, and relax all atomic coordinates until the maximum force is smaller than 0.0001 eV/\AA. The vibrational properties at the $\Gamma$ point are obtained through density-functional perturbation theory \cite{RevModPhys.73.515}, providing the phonon frequencies, $\omega_{\bf Q}$, and eigenvectors (mass-weighted normal modes) $U_{\bf Q}$ \cite{ponceprb}.
Our tight-binding Hamiltonian includes the on-site energy and the ten transfer integrals between nearest and next-nearest neighbors, corresponding to 
 ${\mathbf{R}_m - \mathbf{R}_n } =  0,$ 
${\pm\mathbf{a}, \,}$
${\pm\mathbf{b}, \, }$
${\pm\mathbf{c},  \,}$
${ \pm(\mathbf{a}\pm\mathbf{b}),  \,}$
${\pm(\mathbf{a+c}),  \,}$
${\pm(\mathbf{a-c}),  \,}$
${ \pm(\mathbf{b}\pm\mathbf{c}), \,}$
${  \pm(\mathbf{a}\pm\mathbf{b})/2, \,}$
${   \pm({{\bf a}/2}\pm { {\bf b} /2}+\mathbf{c}), \,}$ and
${   \pm({{\bf a}/2}\pm { {\bf b} /2}- \mathbf{c})}$, where {\bf a}, {\bf b}, and {\bf c} are the lattice vectors.
The electronic eigenvalues provided by this model \cite{ref49coropceanu} are given by:
\begin{align}\label{eq22}
  \varepsilon_{\pm}(\mathbf{k}) 
    = \, & \varepsilon_0
    + 2\varepsilon_a \cos ( \mathbf{k}\cdot\mathbf{a} )
    + 2\varepsilon_b \cos ( \mathbf{k}\cdot\mathbf{b} )
    + 2\varepsilon_c \cos ( \mathbf{k}\cdot\mathbf{c}  ) \nonumber \\
    &  + 2\varepsilon_{ab'} \left[ 
		\cos \left(\mathbf{k}\cdot(\mathbf{a}+\mathbf{b})\right) + 
		\cos \left(\mathbf{k}\cdot(\mathbf{a}-\mathbf{b})\right) \, \right]  \nonumber \\
   & + 2\varepsilon_{ac}  \cos \left( \mathbf{k}\cdot(\mathbf{a}+\mathbf{c}) \right) 
	   + 2\varepsilon_{ac'} \cos \left( \mathbf{k}\cdot(\mathbf{a}-\mathbf{c}) \right)   \nonumber \\
	  &  + 2\varepsilon_{bc'} \left[ \cos \left( \mathbf{k}\cdot(\mathbf{b}+\mathbf{c}) \right)  + 
			                              \cos \left( \mathbf{k}\cdot(\mathbf{b}-\mathbf{c})  \right)  \, \right]  \\
		& \pm 2\varepsilon_{ab} \left[ \cos  \left( \mathbf{k}\cdot\frac{\mathbf{a}+\mathbf{b}}{2} \right) + \cos \left( \mathbf{k}\cdot \frac{\mathbf{a}-\mathbf{b}}{2} \right) \, \right] \nonumber \\
		& \pm 2\varepsilon_{abc} \left[ \cos  \left(\mathbf{k}\cdot\frac{\mathbf{a}+\mathbf{b}+2\mathbf{c}}{2} \right)+ \cos  \left( \mathbf{k}\cdot \frac{\mathbf{a}-\mathbf{b}+2\mathbf{c}}{2}\right)\, \right]  \nonumber \\
		&  \pm 2\varepsilon_{abc'} \left[ \cos  \left(\mathbf{k}\cdot\frac{\mathbf{a}+\mathbf{b}-2\mathbf{c}}{2} \right)+ \cos  \left( \mathbf{k}\cdot \frac{\mathbf{a}-\mathbf{b}-2\mathbf{c}}{2} \right)\,\right] \, . \nonumber 
\end{align}
The $\pm$ signs in Eq. \ref{eq22} arise due to the fact that there are two naphthalene molecules per unit cell \cite{ref49coropceanu}. To obtain the parameters of Eq. \ref{ham11}, we proceed in a manner analogous to the one presented in \cite{PRB1}, though more general. 
The 11 $\varepsilon_{mn}$ parameters used in the tight-binding Hamiltonian are obtained by a least-square fit to the {\it ab initio} band structures (including 264 indepedent $\bf k$-points).
The coupling coefficients $g_{\mathbf{Q}mn}$ and $f_{\mathbf{Q}mn}$ correspond to the first and second derivatives of $\varepsilon_{mn}$, respectively, with respect to structural deformations according to each phonon 
eigenvector, given by
\begin{equation}\label{fgcoefs}
    g_{\nu , mn} = \frac{1}{\hbar \omega_{\nu}} \frac{\delta \varepsilon_{mn}}{\delta U_{\nu}}  \,  ; \quad
    f_{\nu , mn} = \frac{1}{2\hbar \omega_{\nu}} \frac{\delta^2 \varepsilon_{mn}}{\delta U_{\nu}^2} \, .
\end{equation}
We restrict the calculation of coefficients, to $\Gamma$-point phonons considering the nine 
intermolecular phonon branches ($\nu$) with lowest phonon frequencies  \cite{notemodes} as also done in Ref. \cite{PRB1}. For the calculation of the coupling coefficients (Eq. \ref{fgcoefs}), we perform finite-difference displacements of the nuclear positions along the phonon eigenvectors. This results in approximately parabolic curves for each of the tight-binding coefficients as a function of displacement. Each of these curves is fitted to a cubic function, whose slope and curvature are used to evaluate $ g_{\nu , mn}$ and $f_{\nu , mn}$, respectively. We then calculate the coupling coefficients at finite $\bf q$ by assuming the eigenmodes to be dispersionless \cite{PRB1}, i.e.,
$\omega_{{\bf{Q}}} \equiv \omega_{{\bf{q}}\nu} = \omega_{\nu}$,  
$g_{{\bf{Q} } mn} = g_{\nu, mn} (e^{-i{\bf{q}}{\bf{R}}_m} + e^{-i{\bf{q}}{\bf{R}}_n})/2  $, 
$f_{{\bf{Q} } mn} = f_{\nu, mn} (1 + \textrm{cos}({i{\bf{q}}({\bf{R}}_m-{\bf{R}}_n)})/2  $ \cite{SM}.

\begin{center}
\begin{figure}[h]
\includegraphics[width=85mm,scale=0.5]{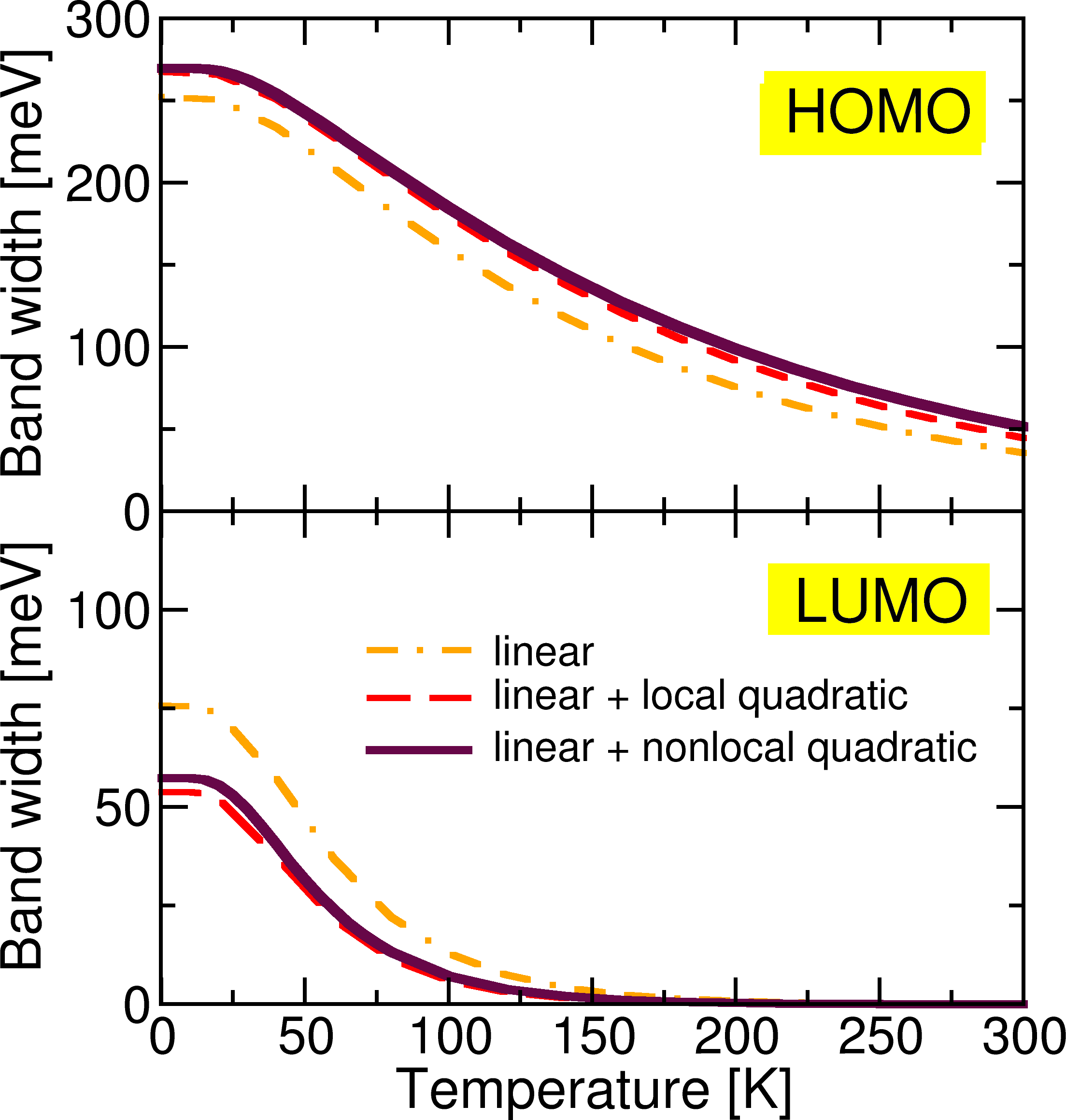}
\caption{\label{naphthalene}{Band widths of the top valence band (HOMO, top) and lowest conduction band (LUMO, bottom) of the naphthalene crystal as a function of temperature. The maroon lines represent results including nonlocal quadratic terms, while red lines include only local quadratic contributions ($f_{{\bf Q}mn} \propto \delta_{mn}$). For comparison, the orange dashed lines corresponds to purely linear electron-vibrational coupling ($f_{\bf Q}=0$ in Eq. \ref{ham11}).} 
}
\end{figure}
\end{center}

In Fig. \ref{naphthalene} we display the band widths of the top valence band and the lowest conduction band (for brevity termed HOMO and LUMO, respectively) as predicted by our new formalism. For comparison, we show the results obtained by purely linear electron-vibrational coupling \cite{PRB1}. The effect of the quadratic coupling gives rise to significant changes, which are a substantial reduction of the LUMO (24\% at T=0) and an increase of the HOMO band width. While the absolute amounts only vary very little along the whole temperature range, the relative changes are significant. For example, the variation of the HOMO band width by second-order coupling is about 45\% at 300K. This is in line with previous works on electron-phonon interaction, which showed that the contribution of the quadratic coupling to band renormalization must not be ignored \cite{ponceprl,giustinoPRL2010,gonze2011}. Here we demonstrate that this also holds for polaron-induced band-narrowing effects. Interestingly, we learn that the quadratic coupling can either increase or decrease the band width, also in line with the effect of the quadratic coupling on band renormalization \cite{gonze2011}. For this system, we find a low contribution of the quadratic nonlocal terms of the Hamiltonian ($f_{{\bf Q} mn }$ with $m \neq n$) \cite{notemodes}.

To summarize this work, we have presented a nonperturbative theory of electron-phonon interaction based on an analytical solution of the quadratic Holstein-Peierls-type Hamiltonian. Despite the inherent complexity of the problem, our results allow for an intuitive interpretation in terms of quasiparticles, i.e., independent polarons and dressed phonons. As all the related material parameters can be obtained from {\it ab initio} calculations, our theory represents an important step towards the quantitative analysis of temperature-dependent band narrowing in real materials beyond perturbational approaches. Moreover, our approach could be extended towards polaron-mobility theories \cite{hanchargetransport,ortmann2011} 
including quadratic electron-phonon scattering processes.

\begin{acknowledgments}
We acknowledge Karsten Hannewald for encouragement on this problem. Financial support from the Helmholtz Energy Alliance of the Helmholtz-Zentrum Berlin and the Deutsche Forschungsgemeinschaft (DFG) - Projektnummer 182087777 - SFB 951 is appreciated. 
\end{acknowledgments}


\begin{thebibliography}{54}
\bibitem[{not({\natexlab{a}})}]{notta1}
  \bibinfo{note}{Consider that we want to calculate the 100 lowest
  eigenvalues of a given Holstein-Peierls Hamiltonian. Taking into account 5
  phonons modes along every Cartesian direction and a population of up to 5
  phonons, with the size of the purely electronic Hamiltonian being 10, would
  require at least 10$^{15}$ floating-point operations compared to 10$^3$
  required for our approach.}
\bibitem[{not({\natexlab{b}})}]{notemodes}
  \bibinfo{note}{The omitted vibrational modes are expected to have only
  small impact on the band narrowing due to their relatively low occupations
  below 300 K \cite{PRB1}. We cannot fully exclude though that some of them
  could have stronger contributions to the nonlocal terms.}
\end{thebibliography}
\end{document}